\documentclass[aps,prl,twocolumn,groupedaddress,showpacs]{revtex4}
\usepackage{graphicx}

\begin{document}
\title{Negative isotope effect of a BCS-like gap;\\
          an inelastic light scattering study}
\author{
In-Sang Yang$^{1}$,
M. V. Klein$^2$, S. Bud'ko$^3$, and
P. C. Canfield$^3$
}

\address{
$^1$Department of Physics, Ewha Womans University, Seoul 120-750, Korea\\
$^2$Materials Research Laboratory, Department of physics,
University of Illinois at Urbana-Champaign, Urbana, IL 61801\\
$^3$Ames Laboratory, U. S. Department of Energy and Department of
Physics and Astronomy, Iowa State University, Ames, IA 50011
}

\date{\today}

\begin{abstract}
In a prior study of the boron isotope effect on the transition
temperature ($T_c$) of
  $R$Ni$_2$B$_2$C ($R$ = Y, Lu) system, a
positive isotope effect was observed, e.g.,
$T_{c}(^{10}B)> T_{c}(^{11}B)$.
BCS theory predicts that the superconducting gap at zero temperature
($\Delta_0$) is proportional to $T_c$.
Therefore, the gap should also
show a positive boron isotope effect.
On the contrary, in an inelastic light scattering study, we report a
negative boron isotope effect on the energy $\Omega$
of the gap-like feature in B$_{2g}$ symmetry, e.g., $\Omega(^{10}B)<
\Omega(^{11}B)$.
The origin of the effect is discussed in terms of
selective probing of a non-conventional electron-phonon interaction
on part of the Fermi surface, probably near the
X point in the Brillouin zone.
\end{abstract}


\maketitle


In a previous study of an isotope effect on the transition
temperature ($T_c$) of $R$Ni$_2$B$_2$C ($R$ = Y, Lu)
system,\cite{cheon} positive isotope exponents $\alpha^{Tc}_{B} =
+ 0.21 \pm 0.07$ for YNi$_2$B$_2$C and $+ 0.11 \pm 0.05$ for
LuNi$_2$B$_2$C were observed upon changing the mass of the boron
site. The exponent $\alpha^{Tc}_{B}$ is defined as
$\alpha^{Tc}_{B} \equiv
   - \frac{\Delta \ln T_c}{\Delta \ln M_B}$.
A relatively large positive value despite the light mass of boron is
regarded as a good indication that
these materials  are BCS-type superconductors, where
phonons involving boron atoms play a major role in the pairing mechanism.

BCS-type theory predicts that the superconducting gap in the limit of the
absolute zero temperature  ($\Delta_0$) is proportional to $T_c$.  In weak
coupling
$\Delta_0 = 1.76 k_B T_c$, where $k_B$ is the Boltzmann constant.
The gap
should also show a positive isotope effect upon changing the  boron mass.
We report here  the contrary result: In an inelastic light scattering
study, we observe a negative isotope effect
of the gap-like Raman peak in $B_{2g}$ symmetry.

The gap $\Delta$ can be measured by electronic Raman scattering.
The photon scattering process excites two quasi-particles out of
the condensate at a minimum  energy cost of $2\Delta$.  In the
limit of small wave-vector transfer $q$ relative to the inverse
coherence length and small elastic and inelastic scattering rate
relative to $2\Delta/\hbar$ and with neglect of final state
interactions, the Raman spectrum has a peak at an energy shift of
$2\Delta$. Earlier measurements on NbSe$_2$ and A15 compounds
showed redistribution of the continuum of the Raman spectra and
the formation of a $2\Delta$-like peak upon cooling the samples
below $T_c$.\cite{a15}
Electronic Raman studies have played
important role in characterizing  the nature of the
superconducting gap of the high-temperaure
superconductors.\cite{cooper}

Compared with other techniques that measure the gap, such as
tunneling and photoemission spectroscopy, inelastic visible light
scattering has the advantage of measuring the anisotropy of the
gap while being less surface sensitive.


The single crystal samples were grown by the flux-
growth method\cite{canfield,xu}
and  were  characterized by  resistivity, magnetization, and neutron
scattering.\cite{character}
$R$Ni$_2$B$_2$C crystalizes in the tetragonal body-centered space group
$I4/mmm$,
and phononic Raman analyses have been made earlier.\cite{p-raman}
Here, we concentrate on the electronic Raman scattering from
the quasi-particles
in normal and superconducting states.

Raman spectra were obtained
in a pseudo-backscattering geometry
using a custom-made subtractive triple-grating spectrometer
designed for very small Raman shifts and ultra low intensities.\cite{kang}
A 3 mW beam of 6471 \AA  \/  Kr-ion laser light was focused onto a spot of
$100 \times 100 \mu$m$^2$.
The temperature rise $\Delta$T of the focal spot above ambient  temperature
was estimated by solving the heat-diffusion equation.
$\Delta$T is largest  at lowest temperature  because of the smaller values
of the thermal conductivity.\cite{thermal}
The estimated $\Delta$T is 2.7K at 4K and 0.9K at 14K for
YNi$_2$B$_2$C single crystals.
The ambient temperature at which the Raman continua begin to show the
redistribution was
determined to be 14K, in agreement with this estimate.
The spectra were corrected for the Bose factor and therefore are
proportional to the
imaginary part of the Raman susceptibility.

The ``$B_{2g}$ spectra'' and ``$B_{1g}$ spectra'' presented below are actually
from the linearly-polarized light scattering geometries of XY
($B_{2g}$ + $A_{2g}$) and
X'Y' ($B_{1g}$ + $A_{2g}$), respectively.
The $A_{2g}$ contribution to these spectra was determined
from the sum of XY  and  X'Y' spectra
minus the spectrum obtained in LR ($B_{1g}$ + $B_{2g}$) circularly-polarized
scattering geometry.  $A_{2g}$ was found to be negligible.


\begin{figure}
\centering
  \includegraphics[width=7.6cm]{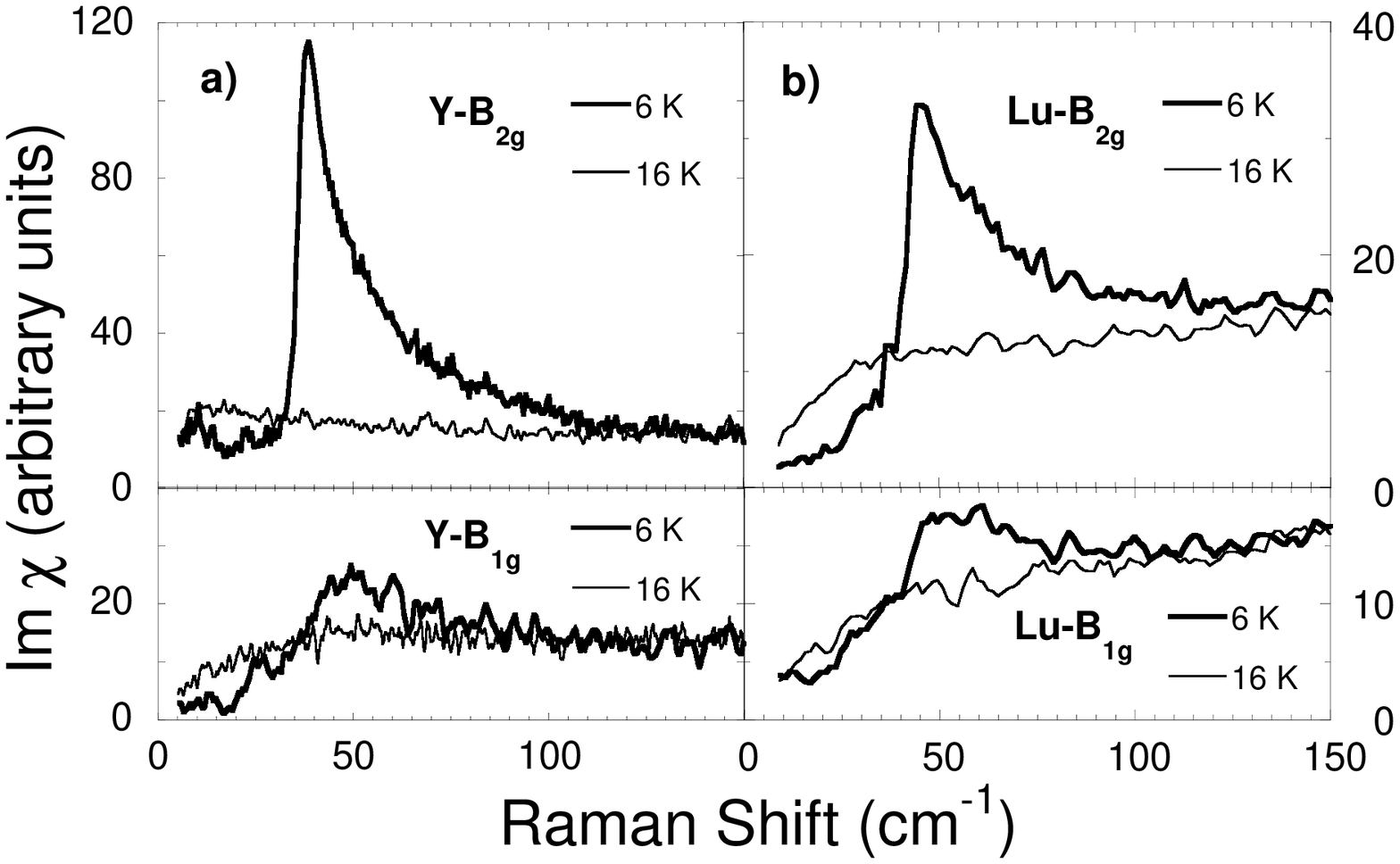}
  \caption{\label{6-16K}
Electronic Raman spectra in $B_{2g}$ and $B_{1g}$ geometries from
(a) YNi$_2$B$_2$C ($^{10}$B) and (b) LuNi$_2$B$_2$C ($^{10}$B) at
6K and 16K (close to $T_c$). Although the Y-axes are in arbitrary
units, the relative intensities are meaningful.
  }
\vspace{1.0cm}
\end{figure}

Figure~\ref{6-16K} shows the electronic Raman spectra in $B_{2g}$
and $B_{1g}$ geometries of (a) YNi$_2$B$_2$C ($^{10}$B) and (b)
LuNi$_2$B$_2$C ($^{10}$B) in the superconducting state (6 K) and
normal state (16 K). The figure clearly shows 2$\Delta$-like peaks
in the superconducting states. As reported earlier,\cite{yang} the
development of the 2$\Delta$-like peaks is indeed due to
superconductivity, following the BCS prediction of the
gap-vs-temperature very well.\cite{other}

The gap values extrapolated to T = 0 are
2$\Delta_0 (B_{2g}$) = 5.0 meV (2$\Delta_0$/kT$_c$ = 3.7) and
2$\Delta_0 (B_{1g}$) = 6.1 meV (2$\Delta_0$/kT$_c$ = 4.5)
  for YNi$_2$B$_2$C,
while
2$\Delta_0 (B_{2g}$) = 6.2 meV (2$\Delta_0$/kT$_c$ = 4.5) and
2$\Delta_0 (B_{1g}$) = 6.6 meV (2$\Delta_0$/kT$_c$ = 4.8)
  for LuNi$_2$B$_2$C.
Finite scattering intensity below the gap-like peaks is observed
even at 6 K. Possibility of signal from the normal-state phase is
ruled out and is discussed in our earlier paper\cite{yang} in more
detail.

We have measured YNi$_2\thinspace^{10}$B$_2$C (B10-Y),
   YNi$_2\thinspace^{11}$B$_2$C (B11-Y),
   LuNi$_2\thinspace^{10}$B$_2$C (B10-Lu), and
   LuNi$_2\thinspace^{11}$B$_2$C (B11-Lu) single crystals
in order to investigate the isotope effect on the ``2$\Delta$-like''
peaks.
Measurements were repeated on 3-4 different single crystals of each set,
and comparisons between the two isotopes were made under conditions that
were as identical as possible.
For instance, one sample of B10-Y and another one of B11-Y were
loaded together into the same cryostat
and measured several times in the sequence of B10 - B11 - B10 - B11 - B10...
in one experimental run.

\begin{figure}
\centering
  \includegraphics[width=7.6cm]{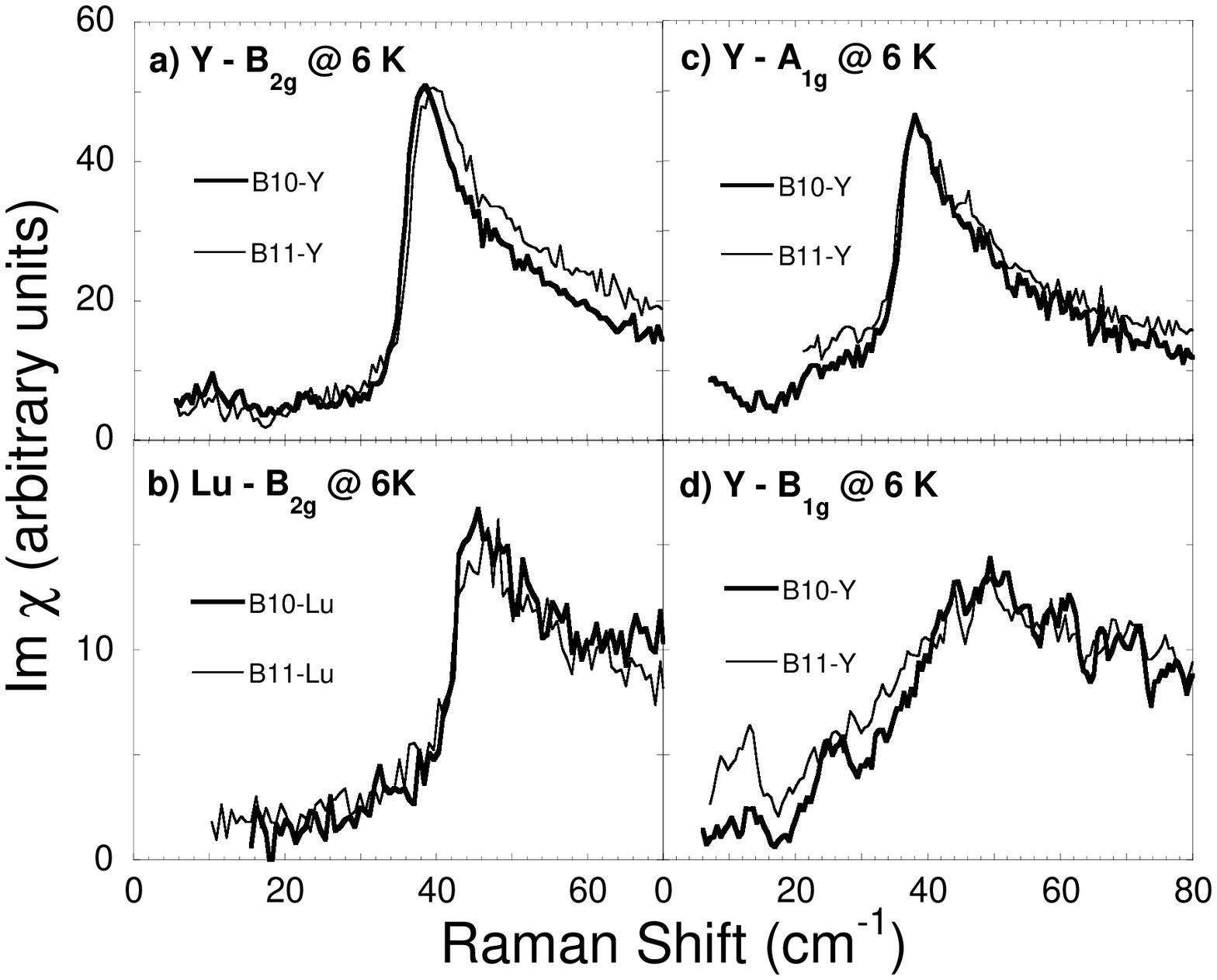}
  \caption{\label{isotope}
Isotope dependance of the the ``2$\Delta$-like'' peaks in Raman
spectra measured at 6 K. a) $B_{2g}$ peaks of B10-Y (thick line)
and B11-Y (thin line), b) $B_{2g}$ peaks of B10-Lu (thick line)
and B11-Lu (thin line), c) $A_{1g}$ peaks of B10-Y (thick line)
and B11-Y (thin line), d) $B_{1g}$ peaks of B10-Y (thick line) and
B11-Y (thin line).
}
\vspace{1.0cm}
\end{figure}

Figure~\ref{isotope} shows the isotope dependence of the
``2$\Delta$-like'' peaks at 6 K in various scattering geometries
for YNi$_2$B$_2$C and $B_{2g}$ geometry for LuNi$_2$B$_2$C.
Clearly, we observe that the $B_{2g}$ peak position in B10-Y is at
{\it smaller} wavenumber than in B11-Y (Fig.~2a). This behavior
was observed in 3-4 pairs of B10-Y and B11-Y. For LuNi$_2$B$_2$C
the trend is far less obvious due to weaker 
signal-to-noise ratio (Fig.~2b). 
In other scattering geometries (Fig.~2c,d), 
no such isotope dependence is observed. On
the contrary, the $B_{1g}$ peak tends to show a positive isotope
effect, but the greater breadth of the peak hinders determining an
isotope effect within experimental resolution.

In our earlier paper,\cite{yang} it was shown that the $B_{2g}$
peak follows the BCS prediction for the temperature dependence of
the superconducting gap. However, as seen in Fig.2a, it does not
follow the BCS prediction for the isotope dependence of the gap.
Rather, the $B_{2g}$ peak frequency in YNi$_2$B$_2$C, as well as
the leading edge of the peak, exhibits a negative isotope effect.
There are prior reports of some of these experimental
results.\cite{prior}

We found empirically that in the region of the $2\Delta$ peaks the
ratio of $B_{2g}$ to $B_{1g}$ Raman intensities can be fit to a
Lorentzian expression
\begin{equation}
\frac{I_{B_{2g}}}{I_{B_{1g}}} =
\frac{A}{(\omega-\omega_0)^2+\Gamma^2}.
\end{equation}
near the `gap-like' peak frequency, $\omega_0$.

\begin{figure}
\centering
  \includegraphics[width=6.0cm]{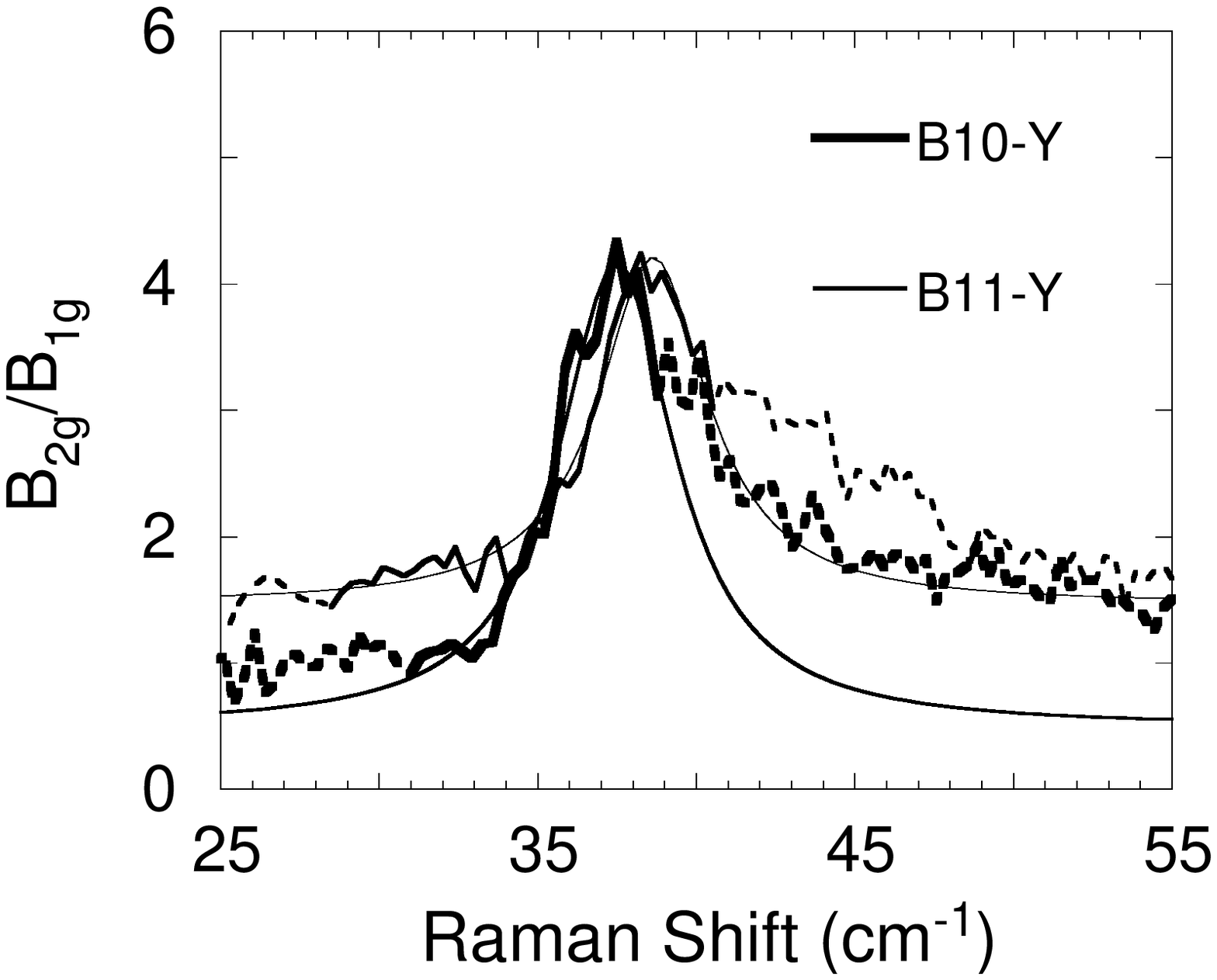}
  \caption{\label{ratio}
Ratios of $B_{2g}$ to $B_{1g}$ intensities and the fit. Fitting
(fine solid lines) of the intensity ratio of the $B_{2g}$ spectra
to the $B_{1g}$ spectra of B10-Y (thick line) and B11-Y (thin
line) in the collective mode scenario. Parts of data which were
fit are shown with solid lines, while data which were excluded
from the fitting are shown with dotted lines. }
\end{figure}

Figure~\ref{ratio} shows the results of fitting the above equation
to the ratio of the $B_{2g}$ to $B_{1g}$ spectra. The fit is
excellent around the peak and below, deviating much at higher
frequencies. No such fit is possible for $I_{B_{2g}}$ or
$I_{B_{1g}}$ itself.
 From the fits to several sets of $B_{2g}$/$B_{1g}$ Raman ratios, we
obtained  the isotope effect exponent
due to boron,
$\alpha^{Raman}_{B} = - 0.32 \pm 0.03$ for YNi$_2$B$_2$C
and $- 0.04 \pm 0.07$ for LuNi$_2$B$_2$C, where
$\alpha^{Raman}_{B} \equiv - \frac{\Delta \ln \omega_0}{\Delta \ln M_B}$.
$M_B$ is the mass of the boron isotopes.  Since there is little evidence
for an isotope effect in the $B_{1g}$ spectra, we attribute these results
for $\alpha^{Raman}_{B}$ primarily to the negative isotope effect on the
$B_{2g}$ spectrum.

\begin{figure}
\centering
  \includegraphics[width=5.5cm]{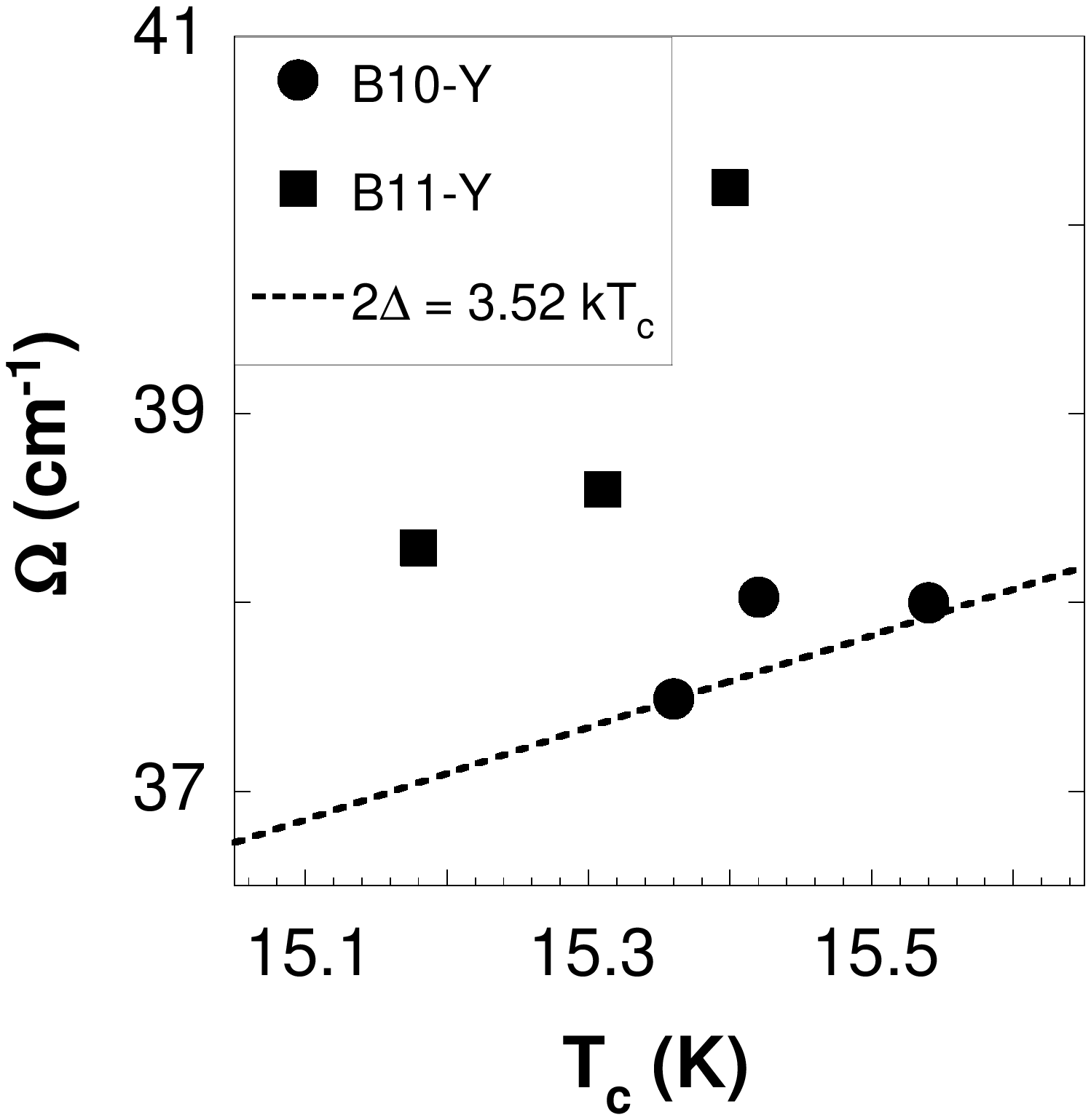}
  \caption{\label{w-T}
Plot of $B_{2g}$ peak frequencies ($\Omega$) at 6K and the onset
temperatures of the superconducting transition of six different
YNi$_2$B$_2$C single crystals. The dashed line on the plot
represents BCS prediction $2\Delta_0 = 3.52 k_B T_c$.}
\end{figure}

We have measured the superconducting transition temperatures using
a SQUID magnetometer and eliminated the possibility that the
particular B10 samples in our study have lower $T_c$ values than
those of B11 samples. Figure~\ref{w-T} shows a plot of $B_{2g}$
peak frequencies ($\Omega$) at 6K and the onset temperatures of
the superconducting transition of six different YNi$_2$B$_2$C
single crystals. As is clearly seen, B11-Y samples do have {\it
larger} $B_{2g}$ peak frequencies than B10-Y samples, in
contradiction to the BCS prediction. This figure shows that the
isotope effect of the $B_{2g}$ peak frequencies is {\it
negatively} correlated with the isotope effect of the $T_c$ in
YNi$_2$B$_2$C. The dashed line on the plot represents $2\Delta_0 =
3.52 k_B T_c$, the BCS prediction in the weak coupling limit.


There is no previous report known to us of an observed negative
isotope effect of the superconducting gap. Negative isotope
effects have been observed for the transition temperatures of the
Pd-H(D) system\cite{pd-h} and  under-doped
YBa$_2$Cu$_3$O$_{7-\delta}$ .\cite{ybco} In the following
discussion we assume that the $B_{2g}$ peak is directly due to
superconducting gap $2\Delta$. How, then, can we explain a
negative isotope effect for the gap measured in $B_{2g}$ Raman
symmetry along with a positive isotope effect for $T_c$ and small
or zero isotope effects for the gap measured in $B_{1g}$ and
$A_{1g}$ symmetries?

The most likely possibility is that the boron portion of the
electron-phonon interaction plays out quite
differently on the various portions of the Fermi surface.  It would
be necessary
for the
$B_{2g}$ Raman vertex to probe those parts of the Fermi surfaces
which have a negative boron-isotope
effect on the value of the gap.    Correspondingly, the isotope
effect would have to be
much smaller and possibly positive on those parts of the Fermi
surface probed in $A_{1g}$ and $B_{1g}$ Raman
geometries.

There is a simple picture of the electronic structure that may give a strong
clue to where the $B_{2g}$ vertex is large.  According to
the band structure calculations for LuNi$_2$B$_2$C\cite{Mattheiss,Pickett}
and YNi$_2$B$_2$C,\cite{Lee} there is a flat band that may
cross the Fermi surface near the X point [wave-vector given by
$(\pi/a,\pi/a,0)$]if there is a small change in parameters.  This
band has strong anti-bonding
$dd\sigma$ interactions as well as $d_{z^2}$ character on Ni
sites.\cite{Pickett}  If we take a nearest neighbor, planar, Ni-Ni
tight binding model with coupling $t$, we obtain a band dispersion
$\bigskip \varepsilon (k)=4t\cos (k_{x}a/2)\cos (k_{y}a/2)$.  This has a
saddle point at X.  A good
starting approximation to the Raman vertex is the so-called mass
approximation, valid when the photon
energy is less than that of all interband transitions starting or
ending at the Fermi energy.  It says that the
$B_{2g}$ Raman vertex is proportional to
$\partial ^{2}\varepsilon /\partial k_{x}\partial k_{y}$.  For the assumed band
structure, this would give $\partial ^{2}\varepsilon /\partial
k_{x}\partial k_{y}=4t(a/2)^{2}\sin(k_{x}a/2)\sin (k_{y}a/2)$, which
takes on its maximum value at the X
point.

It would be useful to verify this suggestion with a band structure
calculation of the mass vertex of the band that
crosses the Fermi energy near the X point.  The next step would be to
study the coupling between this band and the
boron phonon modes.  There is one such calculation in the literature,
which emphasizes the role of the $A_{1g}$
zone center boron mode on modulating the angle of the B-Ni-B
tetrahedra. \cite{MattheissSSC}  However, much
more needs to be done, with a particular look for novel couplings
that would give a negative isotope effect on the
gap associated with this band.

The superconducting gap of borocarbides will not be as simple as
calculated by BCS theory. The finite intensity below the gap
observed in our Raman measurements (Figs. 1 and 2 and\cite{yang}),
a nearly $T^3$ behavior of the electronic specific heat,\cite{cp}
and evidence of small gap on the small pocket Fermi surface from
de Haas-van Alphen measurements\cite{pocket} of superconducting
$R$Ni$_2$B$_2$C ($R$ = Y, Lu) system are good indications of the
complex nature of the gap of borocarbide superconductors.


In conclusion, we observed a negative boron isotope effect of the
gap-like feature in B$_{2g}$ electronic Raman spectra from
YNi$_2$B$_2$C single crystals. For LuNi$_2$B$_2$C, the negative
isotope effect is weaker, and may be zero within our experimental
resolution. A likely explanation is that the $B_{2g}$ Raman vertex
selectively probes those portions of the Fermi surface which show
a negative boron-isotope effect of the order parameter in the
superconducting state.  We have suggested that this may occur at
the X point, the corner of the Brillouin zone in the basal plane.


ISY is grateful for the financial support of the Korea Research
Foundation through a grant number KRF-2000-015-DS0014. MVK was
partially supported under NSF 9705131 and through the STCS under
NSF 9120000. Ames Laboratory is operated by the U.S. Department of
Energy by Iowa State University under Contract No. W-7405-Eng-82.


\bibliography{}

\begin{thebibliography}{99}
\bibitem{other}
No such gap-like peak was observed in $R$Ni$_2$B$_2$C
($R$ = Tm, Er, Ho, Dy) systems, even in the superconducting state.
\end{thebibliography}

\end{document}